\documentclass[aps,prd,twocolumn,preprintnumbers,
superscriptaddress,showpacs,floatfix]{revtex4}

\usepackage{bm}
\usepackage{epsfig}
\usepackage{graphics}
\usepackage{amsmath}

\begin{document}

\preprint{UT-STPD-12/02}

\def\beq{\begin{equation}}
\def\eeq{\end{equation}}
\def\bea{\begin{eqnarray}}
\def\eea{\end{eqnarray}}
\newcommand{\eem}{\end{matrix}}
\newcommand{\bem}{\begin{matrix}}
\newcommand{\beqs}{\begin{subequations}}
\newcommand{\eeqs}{\end{subequations}}
\newcommand{\ftn}{\footnotesize}
\newcommand{\nsz}{\normalsize}
\newcommand{\ssz}{\scriptsize}

\newcommand{\Eref}[1]{Eq.~(\ref{#1})}
\newcommand{\Sref}[1]{Sec.~\ref{#1}}
\newcommand{\Fref}[1]{Fig.~\ref{#1}}
\newcommand{\Tref}[1]{Table~\ref{#1}}
\newcommand{\cref}[1]{Ref.~\cite{#1}}
\newcommand\eqs[2]{Eqs.~(\ref{#1}) and (\ref{#2})}
\newcommand{\tr}{{\mbox{\sf\ssz T}}}
\newcommand\vev[1]{\langle {#1} \rangle}
\def\openep{\leavevmode\hbox{
{\boldmath $\varepsilon$}}}
\def\llgm{\left\lgroup}
\def\rrgm{\right\rgroup}
\newcommand\matt[4]{\mbox{$\llgm\bem #1 &#2 \cr #3& #4\eem\rrgm$}}

\def\lf{\left(}
\def\rg{\right)}
\newcommand{\etal}{{\it et al.\/}}
\newcommand{\GeV}{{\mbox{\rm GeV}}}
\newcommand{\pb}{{\mbox{\rm pb}}}
\def\mcr{{\tt micrOMEGAs}}

\newcommand{\bmm}{{\ensuremath{{\rm BR}\lf B_s\to \mu^+\mu^-\rg}}}
\newcommand{\bsg}{{\ensuremath{{\rm BR}\lf b\to s\gamma\rg}}}
\newcommand{\btn}{{\ensuremath{{\rm R}\lf B_u\to \tau\nu\rg}}}
\newcommand{\Dam}{{\ensuremath{\delta a_{\mu}}}}
\newcommand{\Omx}{{\ensuremath{\Omega_{\rm LSP} h^2}}}
\newcommand{\ssi}{{\ensuremath{\sigma^{\rm SI}_{\tilde\chi p}}}}
\newcommand{\ssd}{{\ensuremath{\sigma^{\rm SD}_{\tilde\chi p}}}}
\newcommand{\mx}{{\ensuremath{m_{\rm LSP}}}}
\newcommand{\Dst}{{\ensuremath{\Delta_{\tilde\tau_2}}}}
\newcommand{\spn}{{\ensuremath{\sigma_{\pi N}}}}
\newcommand{\mst}{{\ensuremath{m_{\tilde\tau_2}}}}
\newcommand{\Mg}{{\ensuremath{M_{1/2}}}}
\newcommand{\AMg}{{\ensuremath{A_0/M_{1/2}}}}
\newcommand{\xx}{{\ensuremath{\tilde\chi}}}
\newcommand{\tnb}{{\ensuremath{\tan\beta}}}
\newcommand{\sign}{{\ensuremath{\rm sign}}}
\newcommand{\Mgut}{\ensuremath{M_{\rm GUT}}}
\newcommand{\Ggut}{\ensuremath{G_{\rm PS}}}
\newcommand{\Gsm}{\ensuremath{G_{\rm SM}}}
\newcommand{\ldt}{\ensuremath{\lambda_{\bf 3}}}
\newcommand{\lds}{\ensuremath{\lambda_{\bf 1}}}
\newcommand{\ld}{\ensuremath{\lambda}}
\newcommand{\TeV}{\ensuremath{\rm TeV}}

\newcommand{\hh}{{\ensuremath{
I{\kern-2.6pt h}}}}
\newcommand{\bhh}{{\ensuremath{\bar{
I{\kern-2.6pt h}}}}}

\renewenvironment{subequations}{%
\refstepcounter{equation}%
\setcounter{parentequation}{\value{equation}}%
  \setcounter{equation}{0}
  \def\theequation{\theparentequation{\sf\small\alph{equation}}}%
  \ignorespaces
}{%
  \setcounter{equation}{\value{parentequation}}%
  \ignorespacesafterend
}

\title{Constrained Minimal Supersymmetric Standard Model \\ with
Generalized Yukawa Quasi-Unification}

\author{N. Karagiannakis}
\email{nikar@auth.gr} \affiliation{Physics Division, School of
Technology, Aristotle University of Thessaloniki, Thessaloniki
54124, Greece}
\author{G. Lazarides}
\email{lazaride@eng.auth.gr} \affiliation{Physics Division, School
of Technology, Aristotle University of Thessaloniki, Thessaloniki
54124, Greece}
\author{C. Pallis}
\email{kpallis@gen.auth.gr} \affiliation{Department of Physics,
University of Cyprus, P.O. Box 20537, CY-1678 Nicosia, CYPRUS}

\date{\today}

\begin{abstract}
We analyze the constrained minimal supersymmetric standard model
with $\mu>0$ supplemented by a generalized `asymptotic' Yukawa
coupling quasi-unification condition, which allows an acceptable
$b$-quark mass. We impose constraints from the cold dark matter
abundance in the universe, $B$ physics, and the mass $m_h$ of the
lightest neutral CP-even Higgs boson. We find that, in contrast to
previous results with a more restrictive Yukawa quasi-unification
condition, the lightest neutralino $\xx$ can act as a cold dark
matter candidate in a relatively wide parameter range. In this
range, the lightest neutralino relic abundance is drastically
reduced mainly by stau-antistau coannihilations and, thus, the
upper bound on this abundance from cold dark matter considerations
becomes compatible with the recent data on the branching ratio of
$B_s\to\mu^+\mu^-$. Also, $m_h\simeq (125-126)~{\rm GeV}$, favored
by LHC, can be easily accommodated. The mass of $\xx$, though,
comes out large ($\sim 1~{\rm TeV}$).
\end{abstract}

 \maketitle

\section{Introduction}\label{sec:intro}

The recently announced experimental data on the mass of the
{\it standard model} (SM)-like Higgs boson \cite{atlas,cms,cdf}
as well as the branching ratio $\bmm$ of the process
$B_s\to\mu^+\mu^-$ \cite{lhcb} in conjunction with
{\it cold dark matter} (CDM)
considerations \cite{wmap} put under considerable stress
\cite{CmssmLhc} the parameter space of the {\it constrained
minimal supersymmetric standard model} (CMSSM)
\cite{Cmssm0, Cmssm,cmssm1,cmssm2}. Let us recall that the CMSSM is a
highly predictive version of the {\it minimal supersymmetric standard
model} (MSSM) based on universal boundary conditions for the soft
supersymmetry (SUSY) breaking parameters. The free parameters of the
CMSSM are
\begin{equation}
\sign\mu,~~\tan\beta,~~\Mg,~~m_0,~~\mbox{and}~~A_0,
\label{param}
\end{equation}
where $\sign\mu$ is the sign of $\mu$, the mass parameter mixing the
electroweak Higgs superfields $H_2$ and $H_1$ of the MSSM which couple
to the up- and down-type quarks respectively, $\tnb$ is the ratio of
the {\it vacuum expectation values} (VEVs) of $H_2$ and $H_1$, and the
remaining symbols above denote the common gaugino mass, the common
scalar mass, and the common trilinear scalar coupling constant,
respectively, defined at the {\it grand unified theory}
(GUT) scale $M_{\rm GUT}$ determined by the unification of the
gauge coupling constants.

It would be interesting to investigate the consequences
of these experimental findings for even more restricted versions of
the CMSSM which can emerge by embedding it in a SUSY GUT model with
a gauge group containing $SU(4)_c$ and $SU(2)_R$. This can lead
\cite{pana} to `asymptotic' {\it Yukawa unification} (YU)
\cite{als}, i.e. the exact unification of the third generation
Yukawa coupling constants (of the top [bottom] quark $h_t$ [$h_b$]
and the tau lepton $h_\tau$) at $M_{\rm GUT}$. The conditions for
this to hold are that the electroweak Higgs superfields $H_2$,
$H_1$ as well as the third generation right handed quark
superfields form $SU(2)_R$ doublets, the third generation
quark and lepton $SU(2)_L$ doublets [singlets] form a $SU(4)_c$
4-plet [$\overline{4}$-plet], and the Higgs doublet $H_1$ which
couples to them is a $SU(4)_c$ singlet. The simplest GUT gauge group
which contains both $SU(4)_{\rm c}$ and $SU(2)_R$ is the
{\it Pati-Salam} (PS) group $G_{\rm PS}=SU(4)_c\times SU(2)_L\times
SU(2)_R$ \cite{leontaris,jean}.

It is well-known that, given the experimental values of the
top-quark and tau-lepton masses (which, combined with YU,
naturally restrict $\tan\beta$ to large values), the CMSSM
supplemented by the assumption of YU yields unacceptable values of
the $b$-quark mass $m_b$ for both signs of $\mu$. This is due to
the generation of sizable SUSY corrections \cite{copw} to $m_b$
(about 20$\%$), which arise from sbottom-gluino (mainly) and
stop-chargino loops \cite{copw, pierce} and have the same sign as
$\mu$ -- with the standard sign convention of Ref.~\cite{sugra}.
The predicted tree-level $m_b(M_Z)$, which turns out to be close
to the upper edge of its $95\%$ {\it confidence level} (c.l.)
experimental range, receives, for $\mu>0$ [$\mu<0$], large
positive [negative] corrections which drive it well above [a
little below] the allowed range. Consequently, for both signs of
$\mu$, YU leads to an unacceptable $m_b(M_Z)$ with the $\mu<0$
case being much less disfavored.

In Ref.~\cite{qcdm} -- see also Refs.~\cite{muneg,nova,yqu,
quasiShafi, pekino} --, concrete SUSY GUT models based on $G_{\rm
PS}$ are constructed which naturally yield a moderate deviation
from exact YU and, thus, can allow acceptable values of the
$b$-quark mass for both signs of $\mu$ within the CMSSM. In
particular, the Higgs sector of the simplest PS model
\cite{leontaris, jean} is extended so that $H_2$ and $H_1$ are not
exclusively contained in a $SU(4)_c$ singlet, $SU(2)_L\times
SU(2)_R$ bidoublet superfield, but receive subdominant
contributions from another bidoublet too which belongs to the
adjoint representation of $SU(4)_c$. As a consequence, a modest
violation of YU is naturally obtained, which can allow acceptable
values of the $b$-quark mass even with universal boundary
conditions. This approach is an alternative to the usual strategy
\cite{raby, baery, shafi, nath} according to which YU is
preserved, but the universal boundary conditions of the CMSSM are
abandoned. We prefer to keep the universality hypothesis for the
soft SUSY breaking rather than the exact Yukawa unification since
we consider this hypothesis as more economical and predictive.
Moreover, it can be easily accommodated within conventional SUSY
GUT models. Indeed, it is known -- cf. first paper in \cref{shafi}
-- that possible violation of universality which could arise from
D-term contributions if the MSSM is embedded into the PS GUT model
does not occur provided that the soft SUSY breaking scalar masses
of the superheavy superfields which break the GUT gauge symmetry
are assumed to be universal.

We will focus, as usually \cite{qcdm,nova,yqu,pekino}, on the
$\mu>0$ case since $\mu<0$ is strongly disfavored by the
constraint arising from the deviation $\delta a_\mu$ of the
measured value of the muon anomalous magnetic moment $a_\mu$ from
its predicted value $a^{\rm SM}_\mu$ in the SM. Indeed, $\mu<0$ is
defended \cite{g2davier} only at $3-\sigma$ by the calculation of
$a^{\rm SM}_\mu$ based on the $\tau$-decay data, whereas there is
a stronger and stronger tendency \cite{Hagiwara, kinoshita} at
present to prefer the $e^+e^-$-annihilation data for the
calculation of $a^{\rm SM}_\mu$, which favor the $\mu>0$ regime.
Note that the results of Ref.~\cite{Jen}, where it is claimed that
the mismatch between the $\tau$- and $e^+e^-$-based calculations
is alleviated, disfavor $\mu<0$ even more strongly.

The representation used for the Higgs superfield which mixes the
$SU(2)_L$ doublets contained in the  $SU(4)_c$ singlet and
non-singlet Higgs bidoublets plays a crucial role in the proposal
of \cref{qcdm}. As argued there, this Higgs superfield can be
either a triplet or a singlet under $SU(2)_R$. In particular, it
was shown that extending the PS model so as to include a pair of
$SU(2)_R$-triplet and/or a pair of $SU(2)_R$-singlet Higgs
superfields belonging to the adjoint representation of $SU(4)_c$
can lead to a sizable violation of YU. However, in the past, we
mainly focused \cite{qcdm,nova,yqu,pekino} on the minimal extension
of the PS model resulting from the inclusion of just a pair of
$SU(2)_R$-triplet superfields since this was enough to generate an
adequate violation of YU ensuring, at the same time, a SUSY
spectrum which leads to successful radiative electroweak symmetry
breaking and a neutralino {\it lightest SUSY particle} (LSP) in a
large fraction of the parametric space. The resulting asymptotic
Yukawa quasi-unification conditions, which replaced the exact YU
conditions, depend only on one new complex parameter ($c$) which
was considered for simplicity real. It is also remarkable
that this model predicts \cite{yqu,pekino} values for the mass
$m_h$ of the CP-even Higgs boson $h$ close to those discovered
\cite{atlas,cms,cdf} by the {\it Large Hadron Collider} (LHC) and
supports new successful versions \cite{axilleas} of the F-term
hybrid inflation based solely on renormalizable superpotential
terms.

However, it has been recently recognized \cite{pekino} that the
lightest neutralino $\xx$ cannot act as a CDM candidate in this
model. This is because the upper bound on the lightest neutralino
relic density from CDM considerations, although this density is
strongly reduced by neutralino-stau coannihilations, yields a very
stringent upper bound on the mass of the lightest neutralino
$m_\xx$, which is incompatible with the lower bound on $m_\xx$
from the data \cite{bmmexp1} on $\bmm$. This result is further
strengthened by the recent measurements \cite{lhcb} on $\bmm$,
which reduce the previous upper bound on this branching ratio and,
thus, enhance even further the resulting lower bound on $m_\xx$.
The main reason for this negative result is that $\tan\beta$
remains large and, thus, the SUSY contribution to $\bmm$, which
originates \cite{bsmm, mahmoudi} from neutral Higgs bosons in
chargino-, $H^\pm$-, and $W^\pm$-mediated penguins and behaves as
$\tan^6\beta/m^4_A$, turns out to be too large ($m_A$ is the mass
of CP-odd Higgs boson). Note, in passing, that even if one
abandons universality in the electroweak Higgs sector and applies
instead the boundary conditions of the so-called \cite{quasiShafi}
NUHM1 model -- with equal soft SUSY breaking masses for $H_1$ and
$H_2$, but different common soft mass $m_0$ for all the other
scalar fields --, $\tnb$ still remains larger than about $55$.
Consequently, even in this case, compatibility of the data on
$\bmm$ \cite{bmmexp1} with the CDM bound on the neutralino relic
density cannot be achieved -- cf. \cref{quasiShafi}.

Therefore, it would be interesting to check if, in the framework
of the CMSSM and consistently with the GUT models of \cref{qcdm},
we can revitalize the candidacy of $\xx$ as a CDM particle,
circumventing the constraint from $\bmm$ and, at the same time,
obtaining experimentally acceptable $m_h$'s. A key-role in
our present investigation is the inclusion of both pairs of
$SU(2)_R$-triplet and singlet Higgs superfields. This allows for a
more general version of the Yukawa quasi-unification conditions --
already extracted in \cref{qcdm} -- which now depend on one real
and two complex parameters. This liberates the third generation
Yukawa coupling constants from the stringent constraint
$h_b/h_t+h_\tau/h_t=2$ obtained in the monoparametric case and,
thus, can accommodate more general values of the ratios
$h_i/h_j$ with $i,j=t,b,\tau$, which are expected, of course, to
be of order unity for natural values of the model parameters.
This allows for lower $\tnb$'s and, consequently, the extracted
$\bmm$ can be reduced to an acceptable level compatible with the
CDM requirement. The allowed parameter space of the model is
then mainly determined by the interplay of the constraints from
$\bmm$ and CDM and the recently announced results of LHC on
the Higgs mass $m_h$.

We first review the salient features of the PS GUT model in
\Sref{theory} and exhibit the cosmological and phenomenological
requirements that we consider in our investigation in
Sec.~\ref{sec:pheno}. We then find the resulting restrictions
on the parameter space of our model and test the perspective
of direct neutralino detectability in Sec.~\ref{results}.
Finally, we summarize our conclusions in Sec.~\ref{con}.

\section{Violating YU within a SUSY PS Model} \label{theory}

The starting point of our construction is the SUSY GUT model
presented in \cref{jean}. It is based on $G_{\rm PS}$, which,
as already mentioned, is the simplest gauge group that can
lead to YU. The representations and transformations under
$G_{\rm PS}$ of the various matter and Higgs superfields of
the model are presented in Table~\ref{tab:fields} ($U_{\rm c}\in
SU(4)_{\rm c}$, $U_{L}\in SU(2)_{L}$, $U_{R}\in SU(2)_{R}$ and
$\tr~,\dagger$, and $\ast$ stand for the transpose, the
hermitian conjugate, and the complex conjugate of a matrix
respectively). The model also possesses a
Peccei-Quinn (PQ) symmetry, a $U(1)$ R symmetry, and a
discrete $Z_{2}^{\rm mp}$ matter parity symmetry with the charges
of the superfields under these extra global symmetries also shown
in Table~\ref{tab:fields}. The matter superfields are $F_i$ and $F^c_i$
($i=1,2,3$), while $H_1$ and $H_2$ belong to the superfield $\hh$.
So, as one can easily see, all the requirements \cite{pana} for
exact YU are fulfilled. The breaking of $G_{\rm PS}$ down to the
SM gauge group $G_{\rm SM}$ is achieved by the superheavy VEVs
($\sim M_{\rm GUT}$) of the
right handed neutrino type components ($\nu^c_H$, $\bar{\nu}^c_H$)
of a conjugate pair of Higgs superfields $H^c$, $\bar{H}^c$. The
model also contains a gauge singlet $S$ which triggers the breaking
of $G_{\rm PS}$, a ${\rm SU(4)}_c$ {\bf 6}-plet $G$ which gives
\cite{leontaris} masses to the right handed down quark type
components of $H^c$, $\bar{H}^c$, and a pair of gauge singlets
$N$, $\bar{N}$ for solving \cite{rsym} the $\mu$ problem of the
MSSM via a PQ symmetry.

In order to allow for a sizable violation of YU, we extend the
model by including three extra pairs of Higgs superfields
$\hh',\bhh'$, $\phi,\bar\phi$, and $\phi',\bar\phi'$, where the
barred superfields are included in order to give superheavy masses
to the unbarred superfields. These extra Higgs superfields
together with their transformation properties and charges are also
included in Table~\ref{tab:fields}. The superfield $\hh'$ belongs
to the ({\bf 15,2,2}) representation of $SU(4)_c$ which is the
only representation, besides ({\bf 1,2,2}), that can couple to the
fermions. On the other hand, $\phi$ and $\phi'$ acquire superheavy
VEVs of order $M_{\rm GUT}$ after the breaking of $G_{\rm PS}$ to
$G_{\rm SM}$. Their couplings with $\bhh^{\prime}$ and $\hh$
naturally generate a ${\rm SU(2)}_R$- and ${\rm
SU(4)}_c$-violating mixing of the $SU(2)_L$ doublets in $\hh$ and
$\hh^{\prime}$ leading, thereby, to a sizable violation of YU.

More explicitly, the part of the superpotential which is relevant
for the breaking of $\Ggut$ to $\Gsm$ is given by
\bea \nonumber W_{\rm H}&=&\kappa S\lf H^c\bar{H}^c-M^2\rg+
m\phi\bar{\phi}+m'\phi'\bar{\phi}'\\&& -S\lf \beta \phi^2+\beta'
\phi^{\prime2}\rg +\lf \lambda\bar{\phi}+\lambda'\bar{\phi}'\rg
H^c\bar{H}^c, \label{superpotential} \eea
where the mass parameters $M,~m$, and $m'$ are of order
$M_{\rm GUT}$, and
$\kappa$, $\beta$, $\beta'$, $\lambda$, and $\lambda'$ are
dimensionless parameters with $M,~m,~m',~\kappa,~
\lambda,~\lambda'>0$ by field redefinitions. For simplicity, we
take $\beta>0$ and $\beta'>0$ (the parameters are normalized
so that they correspond to the couplings between the SM singlet
components of the superfields).

\par
The scalar potential obtained from $W_{\rm H}$ is given by
\begin{eqnarray}
V_{\rm
H}&=&\left\vert\kappa(H^c\bar{H}^c-M^2)-\beta\phi^2-\beta'\phi^{\prime2}
\right\vert^2 \nonumber \\
&& +\left\vert\kappa S+\lambda\bar{\phi}
+\lambda'\bar{\phi}'\right\vert^2\left(\vert
H^c\vert^2+\vert\bar{H}^c \vert^2\right) \nonumber \\
&& +\left\vert 2\beta S\phi-m\bar{\phi} \right\vert^2 +\left\vert
2\beta' S\phi'-m'\bar{\phi}' \right\vert^2\nonumber \\
&& +\left\vert m\phi+\lambda H^c \bar{H}^c\right\vert^2+\left\vert
m'\phi'+\lambda' H^c \bar{H}^c\right\vert^2 \nonumber\\
&&+\ {\rm D-terms}, \label{potential}
\end{eqnarray}
where the complex scalar fields which belong to the SM singlet
components of the superfields are denoted by the same symbols as
the corresponding superfields. Vanishing of the D-terms yields
$\bar{H}^c\,^{*}=e^{i\vartheta}H^c$ ($H^c$, $\bar{H}^c$ lie in the
$\nu^c_H$, $\bar{\nu}^c_H$ direction). We restrict ourselves to
the direction with $\vartheta=0$ which contains the SUSY vacua
(see below). Performing appropriate R and gauge transformations,
we bring $H^c$, $\bar{H}^c$ and $S$ to the positive real axis.

\begin{table}[!t]
\caption{Superfield Content of the Model}
\begin{tabular}{c@{\hspace{0.4cm}}c@{\hspace{0.4cm}}
c@{\hspace{0.4cm}} c@{\hspace{0.4cm}} c@{\hspace{0.4cm}}c}
\toprule
{Super-}&{Represen-}&{Transfor-}&\multicolumn{3}{c}{Global}
\\{fields}&{tations}&{mations}
&\multicolumn{3}{c}{Symmetries}
\\ {}&{under $G_{\rm PS}$}&{under
$G_{\rm PS}$}&{$R$} &{$ PQ$} &{$Z^{\rm mp}_2$}
\\\colrule
\multicolumn{6}{c}{Matter Fields}
\\\colrule
{$F_i$} &{$({\bf 4, 2, 1})$}&$F_iU_L^{\dagger}U^\tr_{\rm c}$&
$1/2$ & $-1$ &$1$
 \\
{$F^c_i$} & {$({\bf \bar 4, 1, 2})$} &$U_{\rm c}^\ast U_R^\ast
F^c_i$&{ $1/2$ }&{$0$}&{$-1$}
\\\colrule
\multicolumn{6}{c}{Higgs Fields}
\\\colrule
{$H^c$} &{$({\bf \bar 4, 1, 2})$}&$U_{\rm c}^\ast U_R^\ast
H^c$&{$0$}&{$0$} & {$0$} \\
{$\bar H^c$}&$({\bf 4, 1, 2})$& $\bar{H}^cU^\tr_R U^\tr_{\rm
c}$&{$0$}&{$0$}&{$0$} \\
{$S$} & {$({\bf 1, 1, 1})$}&$S$&$1$ &$0$ &$0$ \\
{$G$} & {$({\bf 6, 1, 1})$}&$U_{\rm c}GU^\tr_{\rm c}$&$1$ &$0$
&$0$\\\colrule
{$\hh$} & {$({\bf 1, 2, 2})$}&$ U_L\hh U^\tr_R$&$0$ &$1$
&$0$\\\colrule
{$N$} &{$({\bf 1, 1, 1})$}&$N$& {$1/2$}&{$-1$} & {$0$}\\
{$\bar N$}&$({\bf 1, 1, 1})$&$\bar N$& {$0$}&{$1$}&{$0$}
\\\colrule
\multicolumn{6}{c}{Extra Higgs Fields}
\\\colrule
$\hh^{\prime}$&{$({\bf 15, 2, 2})$}&$U_{\rm c}^\ast
U_L\hh^{\prime}U^\tr_RU^\tr_{\rm c}$& $0$ & $1$ &$0$
\\
$\bhh^{\prime}$&{$({\bf 15, 2, 2})$}&$U_{\rm
c}U_L\bhh'U^\tr_RU_{\rm c}^\dagger$& $1$ & $-1$ &$0$
\\\colrule
$\phi$&$({\bf 15, 1, 3})$ &$U_{\rm c}U_R\phi U_R^\dagger U_{\rm
c}^\dagger$& $0$ & $0$ &$0$
\\
$\bar\phi$&{$({\bf 15, 1, 3})$}  &$U_{\rm c}U_R\bar \phi
U_R^\dagger U_{\rm c}^\dagger$& $1$ & $0$ &$0$
\\\colrule
$\phi'$&$({\bf 15, 1, 1})$ &$U_{\rm c}\phi' U_{\rm c}^\dagger$&
$0$ & $0$ &$0$
\\
$\bar\phi'$&{$({\bf 15, 1, 1})$}  &$U_{\rm c}\bar \phi' U_{\rm
c}^\dagger$& $1$ & $0$ &$0$
\\\botrule
\end{tabular}\label{tab:fields}
\vspace*{-13pt}
\end{table}

From the potential in \Eref{potential}, we find that the SUSY
vacuum lies at
\beqs\bea  && \label{vacuum1}\vev{H^c\bar{H}^c}=v^2_0, \\ &&
\label{vacuum11}
\vev{\phi}=v_\phi\left(T_c^{15},1,\frac{\sigma_3}{\sqrt{2}}
\right),\\ && \label{vacuum12}
\vev{\phi'}=v'_\phi\left(T_c^{15},1,\frac{\sigma_{0}}{\sqrt{2}}
\right),\eea and \beq
\vev{S}=\vev{\bar{\phi}}=\vev{\bar{\phi}'}=0, \label{vacuum3} \eeq
where\bea && \lf{\frac{v_0}{
M}}\rg^2=\frac{1}{2\xi}\left(1-\sqrt{1-4\xi}\right),
\\ && v_\phi=-{\ld \frac{v_0^2}{
m}},~~{v'_\phi} =-{\ld' \frac{v_0^2}{ m'}} \label{vacuum2} \eea with
\beq \xi={\frac{M^2}{\kappa}}\lf{\frac{\beta\lambda^2}{
m^2}}+{\frac{\beta'\lambda^{\prime2}}{m^{\prime2}}}\rg<1/4.\eeq \eeqs
The structure of $\vev{\phi}$ and $\vev{\phi'}$ with respect to
(w.r.t.) $G_{\rm PS}$ is shown in Eqs.~(\ref{vacuum11}) and
(\ref{vacuum12}), where
\beqs \bea && T^{15}_c= \frac{1}{2\sqrt{3}}\>{\sf
diag}\lf1,1,1,-3\rg,\\ && \sigma_{3}={\sf diag}\lf1,-1\rg,
~~\mbox{and}~~\sigma_{0}={\sf diag}\lf1,1\rg.\eea\eeqs

The part of the superpotential which is responsible for the
mixing of the doublets in $\hh$ and $\hh'$ is
\beq W_{\rm m}=M_\hh\bhh'\hh^{\prime}+\ldt\phi\bhh'\hh
+\lds\phi'\bhh'\hh,\label{Wm}\eeq
where the mass parameter $M_\hh$ is of order $M_{\rm GUT}$
(made real and positive by field rephasing) and $\ldt$,
$\lds$ are dimensionless complex coupling constants. Note
that the two last terms in the right hand side (RHS) of
\Eref{Wm} overshadow the corresponding ones
from the non-renormalizable $SU(2)_R$-triplet and singlet
couplings originating from the symbolic coupling
$\bar{H}^cH^c\bhh^{\prime}\hh$ (see Ref.~\cite{qcdm}).
Defining properly \cite{qcdm,nova} the relevant couplings
in the RHS of \Eref{Wm}, we obtain the mass terms
\bea\nonumber  W_{\rm m}&=&M_\hh\bar
\hh^{\prime\tr}_1\openep\left(\hh^{\prime}_2+
\alpha_2\hh_2\right)\\
&&+M_\hh\left(\hh^{\prime \tr}_1+\alpha_1
\hh^\tr_1\right)\openep\bhh^{\prime}_2+\cdots,
\label{superheavy}\eea
where $\openep$ is the $2\times 2$ antisymmetric matrix with
$\openep_{12}=1$, the ellipsis includes color non-singlet
components of the superfields, and the complex dimensionless
parameters $\alpha_{1}$ and $\alpha_{2}$ are given by
\beqs\bea\label{alphas1}
\alpha_{1}&=&{\frac{1}{\sqrt{2}M_\hh}}\lf-\lambda_{\bf
3}v_\phi+\lambda_{\bf 1}v'_\phi\rg,
\\ \alpha_{2}&=&{\frac{1}{\sqrt{2}M_\hh}}\lf\lambda_{\bf
3}v_\phi+\lambda_{\bf 1}v'_\phi\rg\cdot \label{alphas2} \eea\eeqs

It is obvious from Eq.~(\ref{superheavy}) that we obtain two
pairs of superheavy doublets with mass $M_\hh$:
\beqs\beq\label{shs1}\bhh^{\prime}_1,~H^{\prime}_2~~\mbox{and}~~
~H^{\prime}_1,~\bhh^{\prime}_2, \eeq where \beq\label{shs11}
H^{\prime}_{r}=\frac{\hh^{\prime}_{r}+\alpha_{r}\hh_{r}}
{\sqrt{1+|\alpha_{r}|^2}},~r=1,2 \eeq\eeqs
(no summation over the repeated index $r$ is implied). The
electroweak doublets $H_r$, which remain massless at the GUT
scale, are orthogonal to the $H^{\prime}_{r}$ directions:
\beq H_r=\frac{-\alpha_r^*\hh^{\prime}_r+\hh_r}
{\sqrt{1+|\alpha_r|^2}} \cdot\label{elws}\eeq
Solving Eqs.~(\ref{shs11}) and (\ref{elws}) w.r.t. $\hh_r$ and
$\hh^\prime_r$, we obtain
\beq
\hh_r=\frac{H_r+\alpha^*_rH^{\prime}_r}{\sqrt{1+|\alpha_r|^2}}
~~\mbox{and}~~\hh^\prime_r=\frac{-\alpha_rH_r+H^{\prime}_r}
{\sqrt{1+|\alpha_r|^2}}\cdot~~~\eeq
The superheavy doublets $H^{\prime}_r$ must have zero VEVs,
which gives
\beq\label{hvev}
\vev{\hh_r}=\frac{\vev{H_r}}{\sqrt{1+|\alpha_r|^2}}
~~\mbox{and}~~\vev{\hh^\prime_r}=\frac{-\alpha_r\vev{H_r}}
{\sqrt{1+|\alpha_r|^2}}\cdot~~~\eeq

The Yukawa interactions of the third family of fermions
are described by the superpotential terms
\beq  W_{\rm Y}=y_{33}F_3\hh
F_3^c+2y'_{33}F_3\hh'F_3^c,
\label{Wy}\eeq
where the factor of two is incorporated in the second term in
the RHS of this equation in order to make $y_{33}^{\prime}$
directly comparable to $y_{33}$, since the doublets in
$\hh^{\prime}$ are proportional to $T^{15}_c$, which is
normalized so that the trace of its square equals unity. From
Eqs.~(\ref{hvev}) and (\ref{Wy}) and using the fact that
$\hh'$ is proportional to $T^{15}_c$ in the $SU(4)_c$ space,
we can readily derive the masses of the third generation
fermions:
\beqs\bea  && \label{mtop}
m_t=\left|\frac{1-\rho\alpha_2/
\sqrt{3}}{(1+|\alpha_2|^2)^{\frac{1}{2}}}y_{33}
v_2\right|,
\\ && \label{mbottom}
m_b=\left|\frac{1-\rho\alpha_1/
\sqrt{3}}{(1+|\alpha_1|^2)^{\frac{1}{2}}}y_{33}
v_1\right|,
\\ && \label{mtau}
m_\tau=\left|\frac{1+\sqrt{3}\rho\alpha_1}
{(1+|\alpha_1|^2)^{\frac{1}{2}}}y_{33}
v_1\right|,
\eea\eeqs
where $\rho\equiv y_{33}^{\prime}/y_{33}$ can be made real
and positive by readjusting the phases of $\hh$, $\hh'$ and
$v_r=\vev{H_r}$. The third generation Yukawa coupling constants
($h_t, h_b$, and $h_\tau$) must then obey the following set of
generalized asymptotic Yukawa quasi-unification conditions:
\bea\nonumber &&
h_t(M_{\rm GUT}):h_b(M_{\rm GUT}):h_\tau(M_{\rm
GUT})= \\ &&
\left|\frac{1-{\rho\alpha_2/\sqrt{3}}}
{\sqrt{1+|\alpha_2|^2}}\right|:
\left|\frac{1-{\rho\alpha_1/\sqrt{3}}}
{\sqrt{1+|\alpha_1|^2}}\right|:
\left|\frac{1+\sqrt{3}\rho\alpha_1}
{\sqrt{1+|\alpha_1|^2}}\right|.~~~~
\label{quasi} \eea
These conditions depend on two complex ($\alpha_1$, $\alpha_2$)
and one real and positive ($\rho$) parameter. For natural values
of $\rho$, $\alpha_1$, and $\alpha_2$, i.e. for values of these
parameters which are of order unity and do not lead to unnaturally
small numerators in the RHS of Eq.~(\ref{quasi}), we expect all
the ratios $h_i/h_j$ with $i,j=t,b,\tau$ to be of order unity. So,
exact YU is naturally broken, but not completely lost since the
ratios of the Yukawa coupling constants remain of order unity
restricting, thereby, $\tnb$ to rather large values. On the other
hand, these ratios do not have to obey any exact relation among
themselves as in the previously studied
\cite{qcdm,muneg,nova,yqu,pekino} monoparametric case. This gives
us an extra freedom which allows us to satisfy all the
phenomenological and cosmological requirements with the lightest
neutralino contributing to CDM.

\section{Cosmological and Phenomenological Constraints}
\label{sec:pheno}

The two-loop renormalization group equations for the Yukawa and
the gauge coupling constants and the one-loop ones for the soft
SUSY breaking parameters are used between the GUT scale
$M_{\rm GUT}$ and a common SUSY threshold $M_{\rm SUSY}\simeq
(m_{\tilde t_1}m_{\tilde t_2})^{1/2}$ ($\tilde t_{1,2}$ are the
stop mass eigenstates), which is determined consistently with
the SUSY spectrum. At $M_{\rm SUSY}$, we impose the conditions
for radiative electroweak symmetry breaking, calculate the SUSY
spectrum employing the publicly available code {\tt SOFTSUSY}
\cite{Softsusy}, and include the SUSY corrections to the
$b$-quark and $\tau$-lepton masses \cite{pierce}. The
corrections to $m_\tau$ (almost 4$\%$) lead \cite{qcdm, muneg}
to a small decrease of $\tan\beta$. The running of the Yukawa
and gauge coupling constants from $M_{\rm SUSY}$ to $M_Z$ is
continued using the SM renormalization group equations.

The pole mass of the top quark is fixed at its central value
$M_t=173~\GeV$ \cite{mtmt}, which corresponds to the running
mass $m_t(m_t) = 164.6~\GeV$. We adopt also the central value
\cite{pdata} of the $\overline{\rm MS}$ $b$-quark mass
$m_b \lf m_b\rg^{\overline{\rm MS}}=4.19~\GeV$, which is
evolved up to $M_Z$ using the central value
$\alpha_s(M_Z)=0.1184$ \cite{pdata} of the strong fine
structure constant and then converted \cite{baermb} to the
$b$-quark mass in the ${\rm \overline{DR}}$ scheme at $M_Z$
yielding $m_b(M_Z)=2.84~\GeV$. Finally, the tau-lepton mass
is taken to be $m_\tau(M_Z) = 1.748~\GeV$.

The model parameters are restricted by a number of
phenomenological and cosmological constraints, which are
evaluated by employing the latest version of the publicly
available code {\tt micrOMEGAs} \cite{micro}. We now briefly
discuss these requirements paying special attention to those
which are most relevant to our investigation.

\paragraph{Cold Dark Matter Considerations.}
\label{phenoa} The $95\%$ c.l. range for the CDM abundance,
according to the results of WMAP \cite{wmap}, is
\beq \Omega_{\rm CDM}h^2=0.1126\pm0.0072. \label{cdmba}\eeq
In the CMSSM, the LSP can be the lightest neutralino $\xx$ and
naturally arises as a CDM candidate. The requirement that its
relic abundance $\Omx$ does not exceed the $95\%$ c.l. upper
bound derived from Eq.~(\ref{cdmba}), i.e.
\beq \Omx\lesssim0.12\label{cdmb},\eeq
strongly restricts the parameter space of the model, since
$\Omx$ generally increases with the mass of the LSP $\mx$
and so an upper bound on $\mx$ can be derived from
Eq.~(\ref{cdmb}). The lower bound on $\Omx$ is not taken into
account in our analysis since other production
mechanisms \cite{scn} of LSPs may be present too and/or other
particles \cite{axino, Baerax} may also contribute to the CDM.
We calculate $\Omx$ using the \mcr\ code, which includes
accurately thermally averaged exact tree-level cross sections
of all the (co)annihilation processes \cite{cmssm1, cdm},
treats poles \cite{cmssm2, qcdm, nra} properly, and uses
one-loop QCD and SUSY QCD corrected \cite{copw, qcdm, microbsg}
Higgs decay widths and couplings to fermions.

\paragraph{The Higgs Boson Mass.}
According to recent independent announcements from the ATLAS
\cite{atlas} and the CMS \cite{cms} experimental teams at the
LHC -- see also \cref{cdf} -- a discovered particle, whose
behavior so far has been consistent with the SM-like Higgs
boson, has mass around $125-126~\GeV$. More precisely the
reported mass is $\lf 126.0 \pm 0.4\pm 0.4\rg~\GeV$ \cite{atlas}
or $\lf 125.3 \pm 0.4 \pm 0.5\rg \GeV$ \cite{cms}. In the
absence of an official combination of these results and allowing
for a theoretical uncertainty of $\pm1.5~\GeV$, we construct a
$2-\sigma$ range adding in quadrature the various experimental
and theoretical uncertainties and taking the upper [lower] bound
from the ATLAS [CMS] results:
\beq 122\lesssim m_h/\GeV\lesssim129.2.\label{mhb} \eeq
This restriction is applied to the mass $m_h$ of the CP-even
Higgs boson $h$ of MSSM. The calculation of $m_h$ in the package
{\tt SOFTSUSY} \cite{Softsusy} includes the full one-loop SUSY
corrections and some zero-momentum two-loop corrections
\cite{2loops}. The results are well tested \cite{comparisons}
against other spectrum calculators.

\paragraph{$B$-Physics Constraints.} We also consider the
following constraints originating from  $B$-meson physics:

\begin{itemize}

\item The branching ratio $\bmm$ of the process
$B_s\to\mu^+\mu^-$ \cite{bsmm,mahmoudi} is to be consistent with
the $95\%$ c.l. bound \cite{lhcb}:
\beq \bmm\lesssim4.2\times10^{-9} \label{bmmb}, \eeq
which is significantly reduced relative to the previous
experimental upper bound \cite{bmmexp1} adopted in \cref{pekino}.
This bound implies a lower bound on $\mx$ since $\bmm$ decreases
as $\mx$ increases. Note that, very recently, the LHCb
collaboration reported \cite{LHCb12} a first evidence for the
decay $B_s\to\mu^+\mu^-$ yielding the following two sided
$95\%$ c.l. bound
\beq 1.1\lesssim\bmm/10^{-9}\lesssim6.4.
\label{bmmb12} \eeq
In spite of this newer experimental upper bound on $\bmm$, we
adopt here the much tighter upper bound on $\bmm$ in \Eref{bmmb}
since we consider it more realistic. As we show below, the upper
bound on the LSP mass $\mx$ which can be inferred from the lower
bound on $\bmm$ in \Eref{bmmb12} does not constrain the parameters
of our model.

\item The branching ratio $\bsg$ of the process $b\to s\gamma$
\cite{microbsg,nlobsg} is to be compatible with the $95\%$ c.l.
range \cite{bsgexp, bsgSM, yqu}:
\beq 2.84\times 10^{-4}\lesssim \bsg \lesssim 4.2\times 10^{-4}.
\label{bsgb} \eeq
Note that the SM plus the $H^\pm$ and SUSY contributions
\cite{microbsg,nlobsg} to $\bsg$ initially increases with $\mx$
and yields a lower bound on $\mx$ from the lower bound in
Eq.~(\ref{bsgb}) -- for higher values of $\mx$, it starts mildly
decreasing.

\item The ratio $\btn$ of the CMSSM to the SM branching ratio of
$B_u\to \tau\nu$ \cite{mahmoudi,Btn} is to be confined in the
$95\%$ c.l. range \cite{bsgexp} :
\beq 0.52\lesssim\btn\lesssim2.04\ .\label{btnb} \eeq
A lower bound on $\mx$ can be derived from the lower bound in this
inequality.

\end{itemize}

\paragraph{Muon Anomalous Magnetic Moment.}
\label{phenoc} The discrepancy $\delta a_\mu$ between the measured
value $a_\mu$ of the muon anomalous magnetic moment and its
predicted value in the SM can be attributed to SUSY contributions
arising from chargino-sneutrino and neutralino-smuon loops. The
relevant calculation is based on the formulas of Ref.~\cite{gmuon}.
The absolute value of the result decreases as
$\mx$ increases and its sign is positive for $\mu>0$. On the other
hand, the calculation of $a^{\rm SM}_\mu$ is not yet stabilized
mainly because of the ambiguities in the calculation of the
hadronic vacuum-polarization contribution. According to the
evaluation of this contribution in Ref.~\cite{g2davier}, there is
still a discrepancy between the findings based on the
$e^+e^-$-annihilation data and the ones based on the $\tau$-decay
data -- however, in \cref{Jen}, it is claimed that this
discrepancy can be alleviated. Taking into account the more
reliable calculation based on the $e^+e^-$ data \cite{Hagiwara},
the recent complete tenth-order QED contribution \cite{kinoshita},
and the experimental measurements \cite{g2exp} of $a_\mu$, we end
up with a $2.9-\sigma$ discrepancy
\beq~\delta a_\mu=\lf24.9\pm8.7\rg\times10^{-10}, \label{g2b1}\eeq
resulting to the following $95\%$ c.l. range:
\beq~7.5\times 10^{-10}\lesssim \delta a_\mu\lesssim 42.3\times
10^{-10}. \label{g2b}\eeq
A lower [upper] bound on $\mx$ can be derived from the upper
[lower] bound in \Eref{g2b}. As it turns out, only the upper bound
on $\mx$ is relevant here. Taking into account the aforementioned
computational instabilities and the fact that a discrepancy at the
level of about $3-\sigma$ cannot firmly establish a real deviation
from the SM value, we restrict ourselves to just mentioning at
which level \Eref{g2b1} is satisfied in the parameter space
allowed by all the other constraints -- cf. \cref{CmssmLhc}.

\section{Restrictions on the SUSY Parameters} \label{results}

Imposing the requirements above, we can delineate the allowed
parameter space. We find that the only constraints
which play a role are the CDM bound in Eq.~(\ref{cdmb}), the
lower bound on $m_h$ in Eq.~(\ref{mhb}), and the bound on $\bmm$
in Eq.~(\ref{bmmb}). In the parameter space allowed by these
requirements, all the other restrictions of Sec.~\ref{sec:pheno}
are automatically satisfied with the exception of the lower
bound on $\delta a_\mu$ in Eq.~(\ref{g2b}). This bound will not
be imposed here as a strict constraint on the parameters of the
model for the reasons explained in Sec.~\ref{sec:pheno}. We
will only discuss at which level \Eref{g2b1} is satisfied in
the parameter space allowed by the other requirements.

In Fig.~\ref{fig:A0tanb}, we present the overall allowed parameter
space in the $tan\beta-A_0/\Mg$ plane. Each point in this shaded
space corresponds to an allowed area in the $M_{1/2}-m_0$ plane
(see below). The lower boundary of the allowed parameter space in
Fig.~\ref{fig:A0tanb} originates from the limit on $\bmm$ in
Eq.~(\ref{bmmb}), except its leftmost part which comes from the
lower bound on $m_h$ in Eq.~(\ref{mhb}) or the CDM bound in
Eq.~(\ref{cdmb}). The upper boundary comes from the CDM bound in
Eq.~(\ref{cdmb}). We see that $\tan\beta$ ranges from about 43.8
to 52. These values are only a little smaller than the ones
obtained for exact YU or the monoparametric Yukawa
quasi-unification conditions discussed in
Refs.~\cite{qcdm,nova,yqu,pekino}. This mild reduction of
$\tan\beta$ is, however, adequate to reduce the extracted $\bmm$
to an acceptable level compatible with the CDM requirement. In the
allowed area of Fig.~\ref{fig:A0tanb}, the parameter $A_0/\Mg$
ranges from about $-3$ to 0.1. We also find that, in this allowed
area, the Higgs mass $m_h$ ranges from 122 to $127.23~{\rm GeV}$
and the LSP mass $m_{\rm LSP}$ from about 746.5 to $1433~{\rm
GeV}$. So we see that, although $m_h$'s favored by LHC can be
easily accommodated, the lightest neutralino mass is large making
its direct detection very difficult. At the maximum allowed $\mx$,
$\bmm$ takes its minimal value in the allowed parameter space.
This value turns out to be about $3.64\times10^{-9}$ and, thus,
the lower bound in \Eref{bmmb12} is satisfied everywhere in the
allowed area in Fig.~\ref{fig:A0tanb}. The range of the
discrepancy $\delta a_\mu$ between the measured muon anomalous
magnetic moment and its SM value in the allowed parameter space of
Fig.~\ref{fig:A0tanb} is about $\lf0.35-2.76 \rg\times10^{-10}$
(note that $\delta a_\mu$ decreases as $\tnb$ or $\Mg$ increases).
Therefore, \Eref{g2b1} is satisfied only at the level of 2.55 to
$2.82-\sigma$. Note that had we considered the $\mu<0$ case,
$\delta a_\mu$ would have been negative and the violation of
\Eref{g2b1} would have certainly been stronger than in the $\mu>0$
case.

\begin{figure}[!t]
\includegraphics[width=85mm,angle=0]{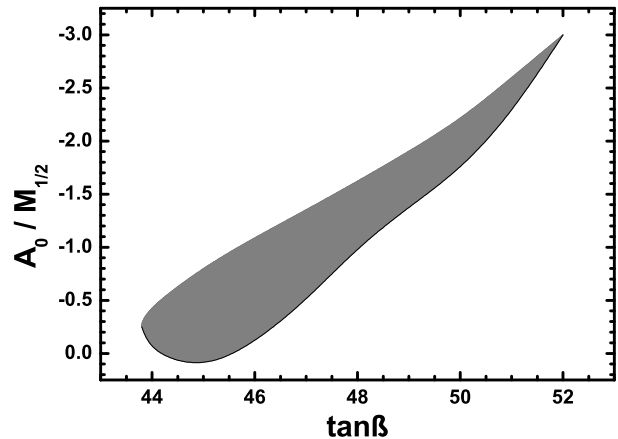}
\caption{The overall (shaded) allowed parameter space of the model
in the $tan\beta-A_0/\Mg$ plane.} \label{fig:A0tanb}
\end{figure}

\begin{figure*}[!t]
\centering
\includegraphics[width=116mm,angle=0]{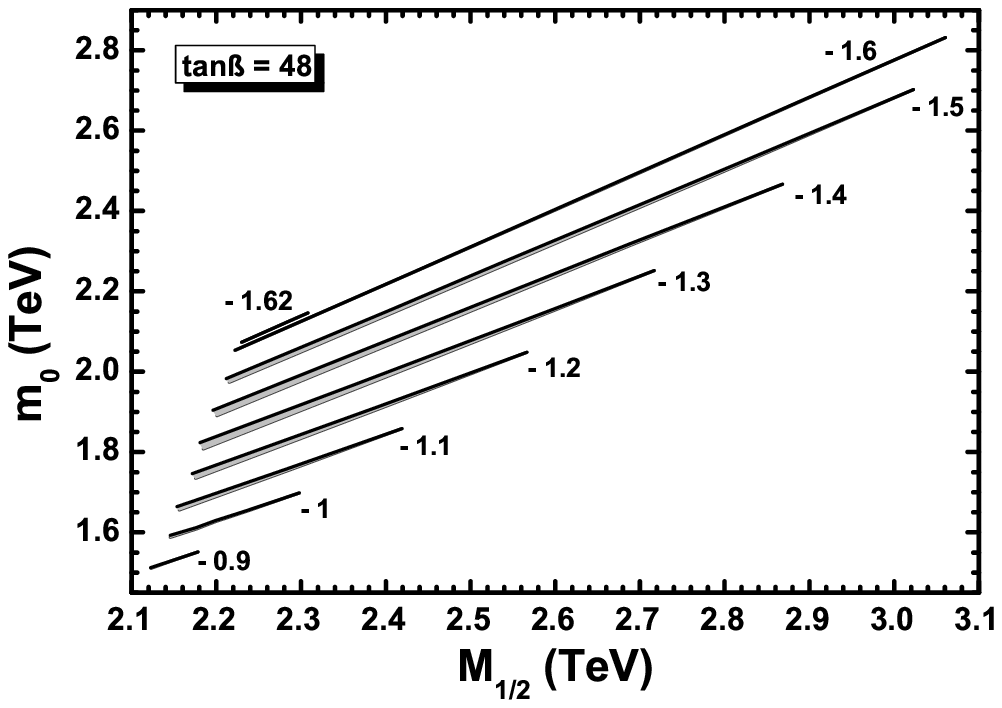}
\caption{The allowed (shaded) areas in the $\Mg-m_0$ plane for
$\tan\beta=48$ and various $A_0/\Mg$'s indicated on the graph.}
\label{fig:tanb48m12m0}
\end{figure*}

\begin{figure*}[!t]
\centering
\includegraphics[width=116mm,angle=0]{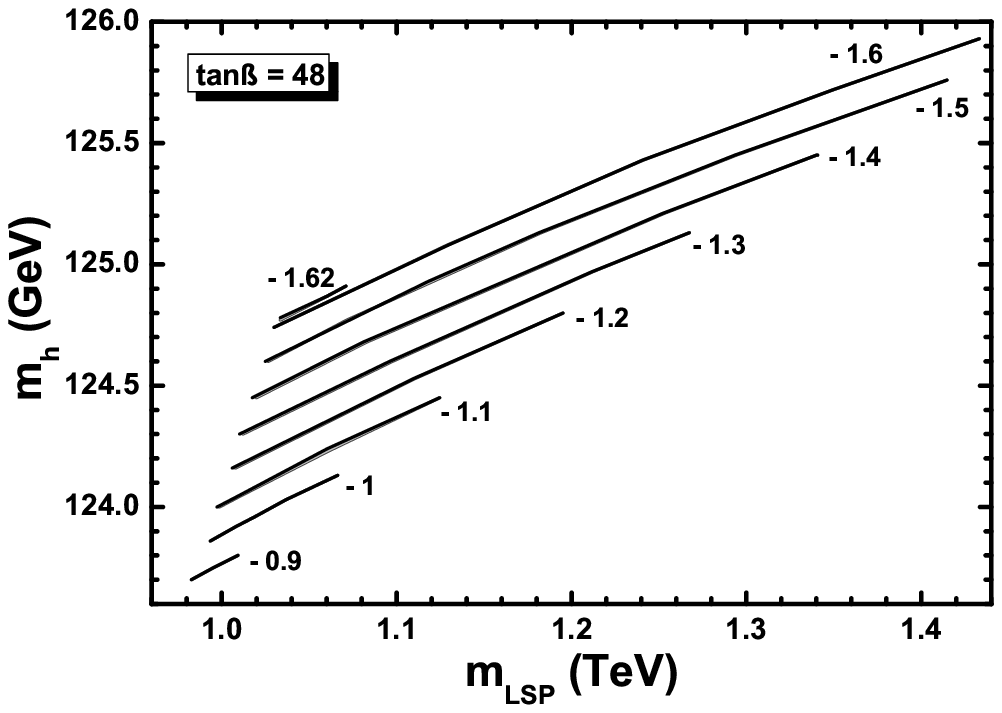}
\caption{The allowed (shaded) areas in the $m_{\rm LSP}-m_h$ plane
for $\tan\beta=48$ and various $A_0/\Mg$'s indicated on the graph.}
\label{fig:tanb48mlspmh}
\end{figure*}

\begin{figure*}[!tb]
\centering
\includegraphics[width=86mm,angle=-0]{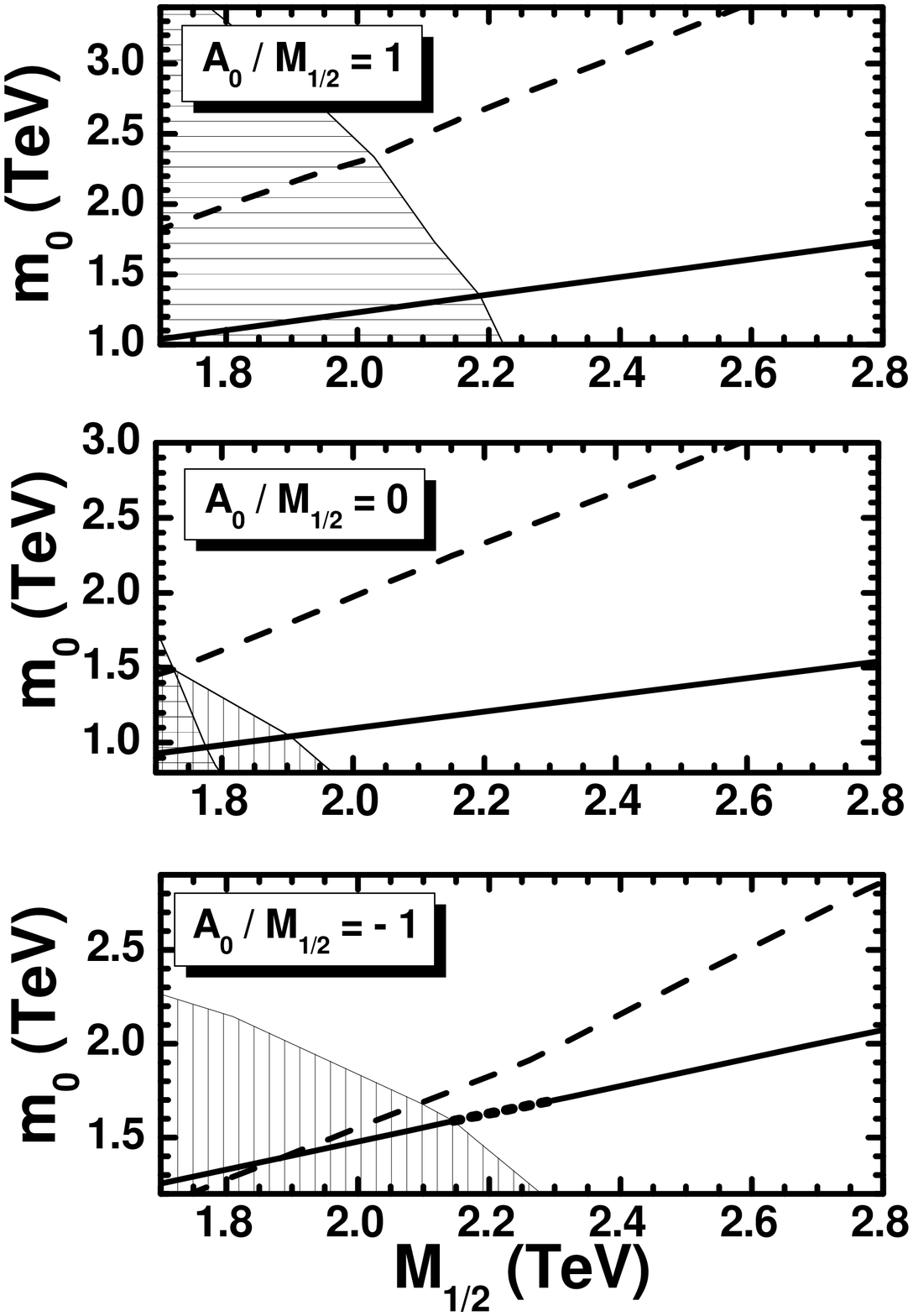}
\includegraphics[width=86mm,angle=-0]{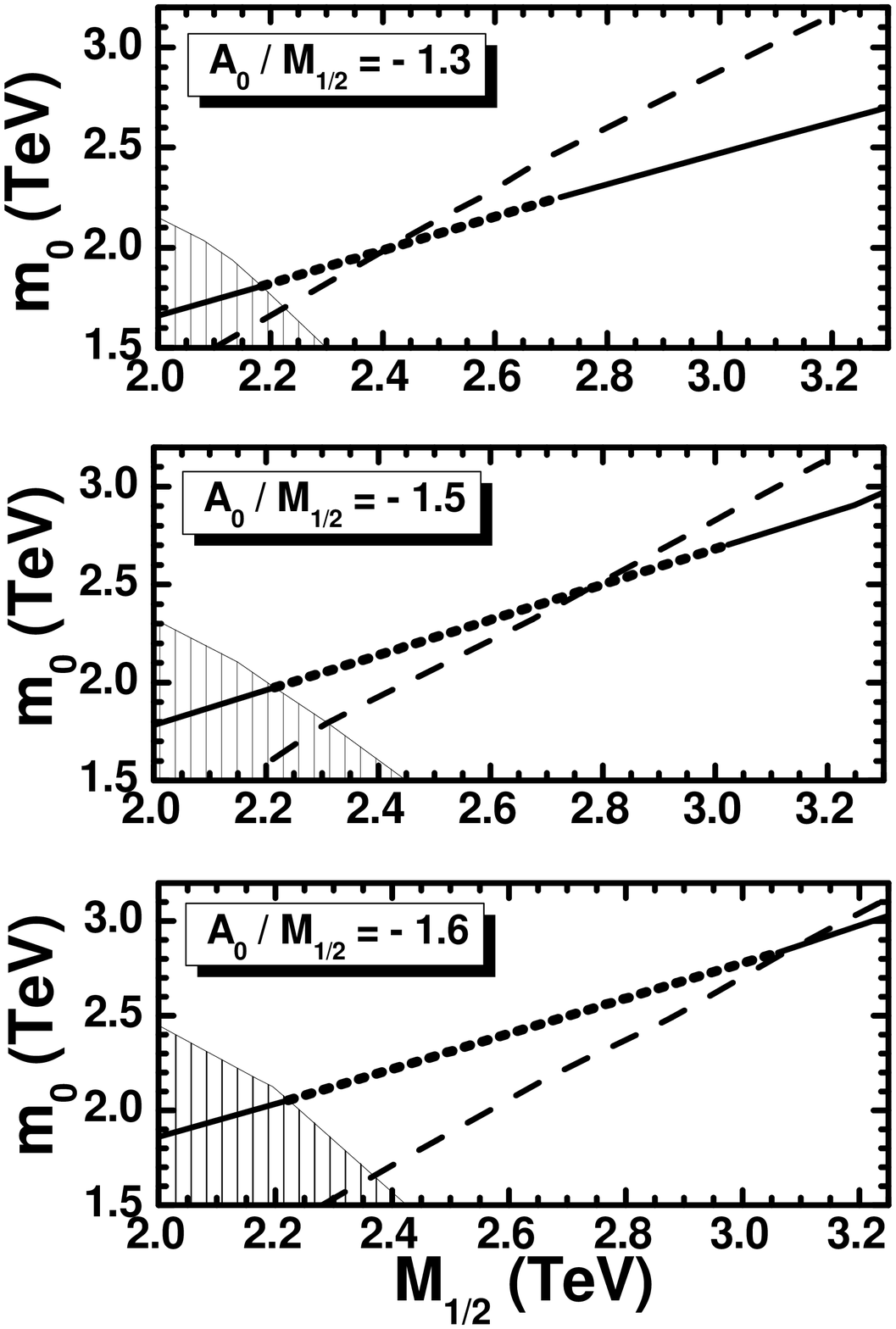}
\caption{Relative position of the $\Dst=0$ (solid) line (with the
dotted part included) and $\Delta_H=0$ (dashed) line for $\tnb=48$
and various $A_0/\Mg$'s indicated on the graphs. Vertically
[horizontally] hatched regions are excluded by the bound in
\Eref{bmmb} [lower bound in \Eref{mhb}]. The dotted areas are the
overall allowed areas.} \label{fig:DAstau}
\end{figure*}

In order to get a better understanding of the structure of the
allowed parameter space and the role played by the various
restrictions, we will now concentrate on the central value of
$\tan\beta=48$ and delineate the allowed areas in the
$M_{1/2}-m_0$ and $m_{\rm LSP}-m_h$ plane for various values of
$A_0/\Mg$. These allowed areas are the shaded areas in
Figs.~\ref{fig:tanb48m12m0} and \ref{fig:tanb48mlspmh}. We observe
that these areas are very thin strips. Their lower boundary
corresponds to $\Dst=0$, where $\Dst=(\mst-\mx)/\mx$ is the
relative mass splitting between the lightest stau mass eigenstate
$\tilde\tau_2$, which is the next-to-LSP, and the LSP. The area
below this boundary is excluded because the LSP is the charged
$\tilde\tau_2$. The upper boundary of the areas comes from the CDM
bound in Eq.~(\ref{cdmb}), while the left one originates from the
limit on $\bmm$ in Eq.~(\ref{bmmb}). The upper right corner of the
areas coincides with the intersection of the lines $\Dst=0$ and
$\Omx=0.12$. We observe that the allowed area, starting from being
just a point at $A_0/\Mg$ slightly bigger than $-0.9$, gradually
expands as $A_0/\Mg$ decreases and reaches its maximal size around
$A_0/\Mg=-1.6$. For smaller $A_0/\Mg$'s, it shrinks very quickly
and disappears just after $A_0/\Mg=-1.62$. We find that, for
$\tan\beta=48$, $m_{\rm LSP}$ ranges from about 983 to $1433~{\rm
GeV}$, while $m_h$ from about 123.7 to $125.93~{\rm GeV}$.

We will now discuss the structure of the allowed areas in
Figs.~\ref{fig:tanb48m12m0} and \ref{fig:tanb48mlspmh}. The fact
that they are narrow strips along the lines with $\Dst=0$
indicates that the main mechanism which reduces $\Omx$ below 0.12
is the coannihilation of $\tilde\tau_2$'s and $\xx$'s. Indeed, we
find that the dominant processes are the
$\tilde\tau_2\tilde\tau_2^*$ coannihilations to $b\bar{b}$ and
$\tau\bar\tau$ contributing to the inverse of $\Omx$ about
$(55-72)\%$ and $(11-15)\%$ respectively. As already noticed in
\cref{CmssmLhc}, these processes are enhanced by the $s$-channel
exchange of the heavy CP-even neutral Higgs boson $H$, with mass
$m_H$, in the presence of a resonance ($2\mx\simeq m_H$)
 -- for the relevant channels, see, for example, Ref.~\cite{cdm}.
In order to pinpoint the effect of the $H$-pole on the
$\tilde\tau_2\tilde\tau_2^*$ coannihilations, we must track its
position relative to the line $\Dst=0$. In Fig.~\ref{fig:DAstau},
the dashed lines correspond to the $H$-pole, i.e. to $\Delta_H=0$
with $\Delta_H=(m_H-2\mx)/2\mx$ for $\tan\beta=48$ and various
values of $A_0/M_{1/2}$ as indicated. The solid lines (with their
dotted part included) correspond to $\Dst=0$ and the vertically
[horizontally] hatched regions are excluded by the bound on $\bmm$
in \Eref{bmmb} [lower bound on $m_h$ in \Eref{mhb}]. We observe
that, for $\AMg = 1$, the lower bound on $\Mg$ which originates
from the lower bound on $m_h$ in \Eref{mhb} overshadows the one
from \Eref{bmmb}. In all other cases, however, we have the
opposite situation. This is consistent with the fact that for
almost fixed $\Mg$ and $m_0$, the Higgs mass $m_h$ increases as
$\AMg$ decreases -- cf.~\cref{CmssmLhc}.

From Fig.~\ref{fig:DAstau}, we see that, for $A_0/M_{1/2}=1$ and
0, the $H$-pole line is far from the part of the $\Dst=0$ line
allowed by all the other constraints without considering the CDM
bound. Consequently, in the neighborhood of this part, the effect
of the $H$-pole is not strong enough to reduce $\Omx$ below 0.12
via $\tilde\tau_2\tilde\tau_2^*$ coannihilations and no overall
allowed area exists. On the contrary, for $A_0/M_{1/2}=-1$, the
$H$-pole line gets near the otherwise allowed (i.e. allowed by all
the other requirements without considering the CDM bound) part of
the $\Dst=0$ line and starts affecting the neighborhood of its
leftmost segment, where $\Omx$ becomes smaller than 0.12 and,
thus,  an overall allowed (dotted) area appears. For
$A_0/M_{1/2}=-1.3$, $-1.5$, $-1.6$, the $H$-pole line moves
downwards and intersects the $\Dst=0$ line with the point of
intersection moving to the right as $A_0/M_{1/2}$ decreases. This
enhances $H$-pole $\tilde\tau_2\tilde\tau_2^*$ coannihilation in
the neighborhood of a bigger and bigger segment of the otherwise
allowed part of the $\Dst=0$ line and, thus, leads to $\Omx$'s
below 0.12 generating an overall allowed (dotted) area. For even
smaller $A_0/M_{1/2}$'s, the $H$-pole line keeps moving downwards
and gets away from most of the otherwise allowed part of the
$\Dst=0$ line. Also, the intersection of these two lines moves to
higher values of $M_{1/2}$ and $m_0$ and the effect of the
$H$-pole is weakened even around this intersection. So the overall
allowed area quickly disappears as $A_0/M_{1/2}$ moves below
$-1.6$.

\begin{figure}[!t]
\includegraphics[width=65mm,angle=-90]{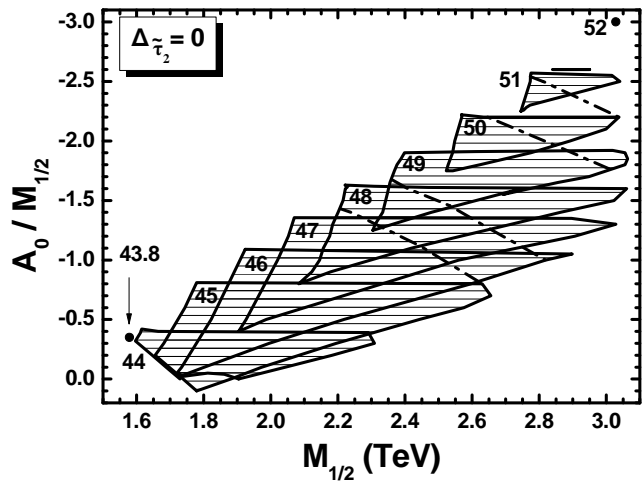}
\caption{Allowed regions in the $M_{1/2}-A_{0}/M_{1/2}$ plane for
$\Delta_{\tilde\tau_2}=0$ and various $\tnb$'s indicated on the
graph. The dot-dashed lines from top to bottom correspond to
$m_h=126.5$, 126, 125, $124.5~{\rm GeV}$.} \label{fig:AMgx}
\end{figure}

\begin{table}[!t]
\caption{Input and output parameters, masses of the sparticles and
Higgses and values of the low energy observables of our model in
four cases (recall that $1~\pb\simeq2.6\times10^{-9}~\GeV^{-2}$).}
\vspace*{1.1mm}
\begin{tabular}{c@{\hspace{0.3cm}}c@{\hspace{0.6cm}}c@
{\hspace{0.6cm}}c@{\hspace{0.6cm}}c} \toprule
\multicolumn{5}{c}{Input Parameters}\\\colrule
$\tan\beta$ & $48$ & $49$ & $50$ & $51$\\
$-A_0/\Mg$ &$1.4$ &$1.6$&$2$ &$2.5$ \\
$\Mg/\TeV$ & $2.27$ &$2.411$&$2.824$ &$2.808$ \\
$m_0/\TeV$ &$1.92$ &$2.295$&$3.156$&$3.747$ \\ \colrule
\multicolumn{5}{c}{Output Parameters}\\\colrule
$h_t/h_\tau(M_{\rm GUT})$ & $1.117$ & $1.079$ & $1.038$ & $1.008$\\
$h_b/h_\tau(M_{\rm GUT})$ & $0.623$ & $0.618$ & $0.613$ & $0.607$\\
$h_t/h_b(M_{\rm GUT})$ & $1.792$ & $1.745$ & $1.693$ &
$1.660$\\\colrule
$\mu/\TeV$ & $2.78$ & $3.092$ & $3.823$ & $4.129$\\
$\Dst (\%)$ & $1.43$ & $0.93$ & $0.1$ & $0.17$\\
$\Delta_H (\%)$ & $3.08$ & $1.30$ & $0.11$ & $1.76$\\
\colrule
\multicolumn{5}{c}{Masses in ${\rm TeV}$ of Sparticles and
Higgses}\\\colrule
$\tilde\chi$& $1.023$ &$1.110$ &$1.309$ &$1.303$\\
$\tilde\chi_2^{0}$ &$1.952$ &$2.117$ &$2.489$ &$2.481$\\
$\tilde{\chi}_{3}^{0}$ &$2.782$ &$3.088$ &$3.815$ &$4.114$\\
$\tilde{\chi}_{4}^{0}$ &$2.785$ &$3.091$ &$3.817$ &$4.116$\\
$\tilde{\chi}_{1}^{\pm}$ &$1.985$ &$2.117$ &$2.489$ &$2.481$\\
$\tilde{\chi}_{2}^{\pm}$ &$2.785$ &$3.091$ &$3.817$ &$4.116$ \\
$\tilde{g}$&$4.809$ &$5.190$ &$6.042$ &$6.040$ \\ \colrule
%
$\tilde{t}_1$ &$3.806$ &$4.097$ &$4.761$ &$4.781$\\
$\tilde{t}_2$ &$3.226$ &$3.458$ &$3.967$ &$3.902$ \\
$\tilde{b}_1$ &$3.838$ &$4.141$ &$4.853$ &$4.947$ \\
$\tilde{b}_2$ &$3.763$ &$4.058$ &$4.733$ &$4.757$\\
$\tilde{u}_{L}$&$4.687$ &$5.138$ &$6.186$ &$6.483$\\
$\tilde{u}_{R}$ &$4.485$ &$4.923$ &$5.946$ &$6.257$ \\
$\tilde{d}_{L}$& $4.687$ &$5.138$ &$6.187$ &$6.483$\\
$\tilde{d}_{R}$&$4.459$ &$4.896$ &$5.914$ &$6.227$\\\colrule
$\tilde\tau_1$&$2.082$ &$2.347$ &$2.979$ &$3.293$\\
$\tilde\tau_2$&$1.037$ &$1.121$ &$1.310$ &$1.305$\\
$\tilde\nu_\tau$ &$2.075$ &$2.342$ &$2.975$ &$3.289$\\
$\tilde{e}_L$&$2.453$ &$2.819$ &$3.690$ &$4.201$\\
$\tilde{e}_R$&$2.112$ &$2.476$ &$3.339$ &$3.901$\\
$\tilde{\nu}_{e}$ &$2.451$ &$2.818$ &$3.689$ &$4.200$\\\colrule
%
$h$&$0.1245$ &$0.125$ &$0.126$&$0.1265$\\
$H$&$2.109$  &$2.249$ &$2.621$&$2.652$\\
$H^{\pm}$&$2.111$  &$2.251$ &$2.623$&$2.654$ \\
$A$&$2.110$ & $2.25$&$2.622$ &$2.652$\\\colrule
\multicolumn{5}{c}{Low Energy Observables}\\\colrule
%
$10^4\bsg$ &$3.25$  &$3.25$ &$3.26$&$3.26$\\
$10^9\bmm$   &$4.17$ &$4.15$ &$3.98$&$4.17$\\
$\btn$   &$0.975$  &$0.977$ &$0.982$&$0.982$\\
$10^{10}\Dam$  &$1.11$&$0.89$ &$0.57$ &$0.49$\\\colrule
$\Omx$ &$0.11$&$0.11$&$0.11$&$0.11$\\
$\ssi / 10^{-12} \pb$ &$6.17$&$4.55$ &$2.44$&$1.75$\\[1mm]
$\ssd / 10^{-9} \pb$ &$1.69$&$1.08$ &$0.43$&$0.28$\\[1mm]\botrule
\end{tabular}
\label{tab:spectrum}
\end{table}

\begin{figure*}[!t]
\centering
\includegraphics[width=89mm,angle=-0]{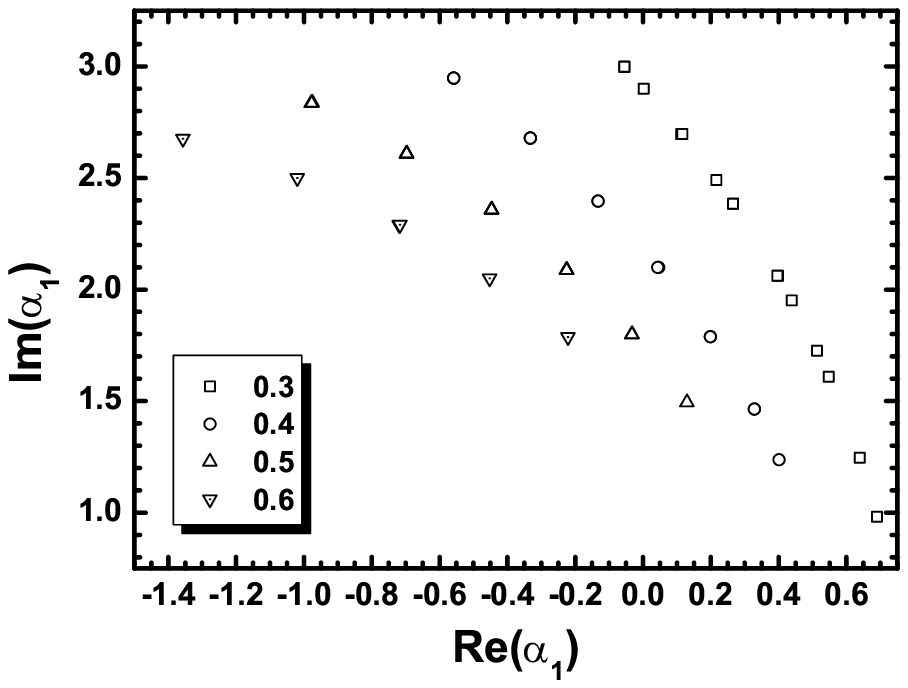}
\includegraphics[width=89mm,angle=-0]{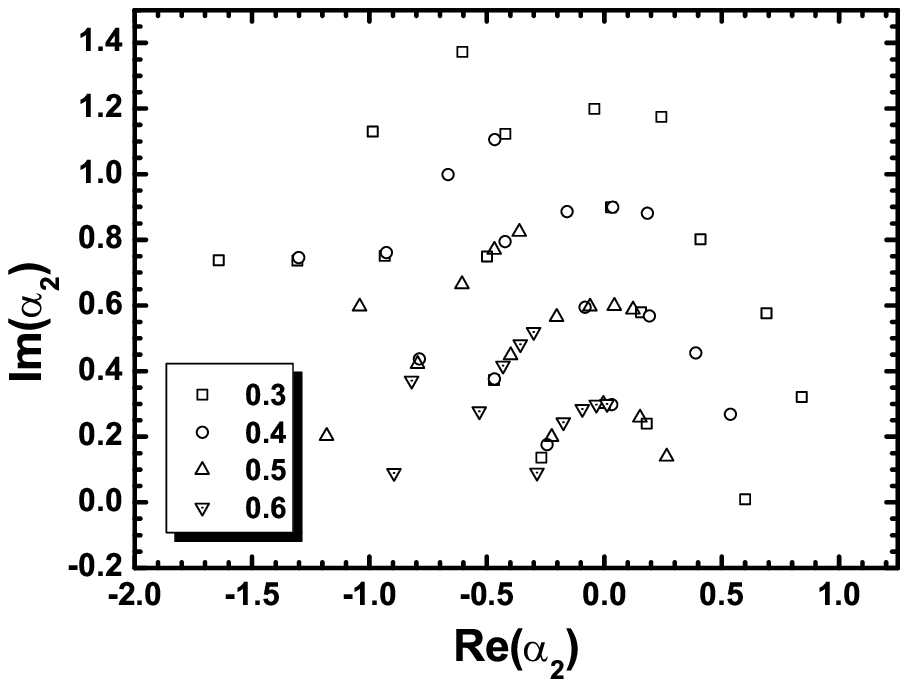}
\caption{The complex parameters $\alpha_1$ and $\alpha_2$ for
various $\rho$'s indicated on the graphs for the case in the
second column of Table~\ref{tab:spectrum}.} \label{fig:a1a2}
\end{figure*}

As we have seen, in the allowed parameter space of our model,
$\Dst$ is very close to zero. So we can restrict ourselves to
$\Dst=0$ without much loss. Note, by the way, that this choice
ensures the maximal possible reduction of $\Omx$ due to
$\tilde\tau_2\tilde\tau_2^*$ coannihilations and, thus, leads to
the maximal allowed $\Mg$ [or $\mx$] for given $\tan\beta$,
$A_0/\Mg$, and $m_0$ [or $m_h$]. In Fig.~\ref{fig:AMgx}, we
present the allowed areas in the $M_{1/2}-A_0/M_{1/2}$ plane for
$\Dst=0$ and various values of $\tan\beta$ indicated on the graph.
They are the horizontally hatched regions. Their right boundaries
correspond to $\Omx=0.12$, while the left ones saturate the bound
on $\bmm$ in Eq.~(\ref{bmmb}) -- cf. \Fref{fig:A0tanb}. The almost
horizontal upper boundaries correspond to the sudden shrinking of
the allowed areas which, as already discussed, is due to the
weakening of the $H$-pole effect as $A_0/M_{1/2}$ drops below a
certain value for each $\tan\beta$. The lower left boundary of the
areas for $\tan\beta=44$, 45, and 46 comes for the lower bound on
$m_h$ in Eq.~(\ref{mhb}), while the somewhat curved almost
horizontal part of the lower boundary of the area for
$\tan\beta=44$ originates from the CDM bound in Eq.~(\ref{cdmb}).
The dot-dashed lines from top to bottom correspond to $m_h=126.5$,
126, 125, $124.5~{\rm GeV}$. We see that the $m_h$'s which are
favored by LHC can be readily obtained in our model for the higher
allowed values of $\tan\beta$.

In Table~\ref{tab:spectrum}, we list the input and the output
parameters of the present model, the masses in $\TeV$ of the
SUSY particles -- gauginos
$\tilde\chi,~\tilde\chi_2^{0},~\tilde{\chi}_{3}^{0},
~\tilde{\chi}_{4}^{0}, ~\tilde{\chi}_{1}^{\pm}~
\tilde{\chi}_{2}^{\pm},~\tilde{g}$, squarks $\tilde{t}_1,~
\tilde{t}_2,~\tilde{b}_1,~\tilde{b}_2,~\tilde{u}_{L},~
\tilde{u}_{R},~\tilde{d}_{L},~\tilde{d}_{R}$, and sleptons
$\tilde\tau_1,~\tilde\tau_2,\tilde\nu_\tau,~\tilde{e}_L,~
\tilde{e}_R,~\tilde{\nu}_{e}$ -- and  Higgses ($h,~H,~H^\pm,~A$)
and the values of the various low energy observables in four
characteristic cases. Note that we consider the squarks and
sleptons of the two first generations as degenerate. From the
values of the various observable quantities it is easy to verify
that all the relevant constraints are met. In the low energy
observables, we included the spin-independent (SI) and
spin-dependent (SD) lightest neutralino-proton ($\xx-p$)
scattering cross sections $\ssi$ and $\ssd$, respectively, using
central values for the hadronic inputs -- for the details of the
calculation, see Ref.~\cite{yqu}. We see that these cross sections
are well below not only the present experimental upper bounds, but
even the projected sensitivity of all planned future experiments.
So the allowed parameter space of our model will not be accessible
to the planned CDM direct detection experiments based on
neutralino-proton scattering. We also notice that the sparticles
turn out to be very heavy, which makes their discovery a very
difficult task.

In the overall allowed parameter space of our model in
Fig.~\ref{fig:A0tanb}, we find the following ranges for the ratios
of the asymptotic third generation Yukawa coupling constants:
$h_t/h_\tau\simeq 0.98-1.29$, $h_b/h_\tau\simeq 0.60-0.65$, and
$h_t/h_b\simeq 1.62-2.00$. We observe that, although exact YU is
broken, these ratios remain close to unity. They can generally be
obtained by natural values of the real and positive parameter
$\rho$ and the complex parameters $\alpha_1$, $\alpha_2$, which
enter the Yukawa quasi-unification conditions in
Eq.~(\ref{quasi}). Comparing these ratios with the ones of the
gauge coupling constants of the non-SUSY SM at a scale close to
$\Mgut$ -- see e.g. \cref{book} --, we can infer that the ratios
here are not as close to unity. In spite of this, we apply the
term Yukawa quasi-unification in the sense that the ratios of the
Yukawa coupling constants in our model are much closer to unity
than in generic models with lower values of $\tan\beta$ -- cf.
\cref{Antuch}. Finally, note that the deviation from exact YU here
is comparable to the one obtained in the monoparametric case --
cf. \cref{yqu} -- and is also generated in a natural, systematic,
controlled, and well-motivated manner.

In order to see this, we take as a characteristic example the
second out of the four cases presented in
Table~\ref{tab:spectrum}, which yields $m_h=125~{\rm GeV}$ favored
by the LHC. In this case, where $h_b/h_\tau=0.618$ and
$h_t/h_\tau=1.079$, we solve Eq.~(\ref{quasi}) w.r.t. the complex
parameters $\alpha_1$, $\alpha_2$ for various values of the real
and positive parameter $\rho$. Needless to say that one can find
infinitely many solutions since we have only two equations and
five real unknowns. Some of these solutions are shown
Fig.~\ref{fig:a1a2}. Note that the equation for $h_b/h_\tau$
depends only on the combination $\rho\alpha_1$ and, thus, its
solutions are expected to lie on a certain curve in the complex
plane of this combination. Consequently, in the $\alpha_1$ complex
plane, the solutions should be distributed on a set of similar
curves corresponding to the various values of $\rho$. This is
indeed the case as one can see from the left panel of
Fig.~\ref{fig:a1a2}. For each $\alpha_1$ and $\rho$ in this panel,
we then solve the equation for $h_t/h_\tau$ to find $\alpha_2$. In
the right panel of Fig.~\ref{fig:a1a2}, we show several such
solutions. Observe that the equation for $h_t/h_\tau$ depends
separately on $\alpha_2$ and $\rho$ and, thus, its solutions do
not follow any specific pattern in the $\alpha_2$ complex plane.
Note that each point in the $\alpha_1$ plane generally corresponds
to more than one points in the $\alpha_2$ plane. We scanned the
range of $\rho$ from 0.3 to 3 and found solutions only for the
lower values of this parameter (up to about 0.6). The solutions
found for $\alpha_1$ and $\alpha_2$ are also limited in certain
natural regions of the corresponding complex planes. The picture
is very similar to the one just described for all the possible
values of the ratios of the third generation Yukawa coupling
constants encountered in our investigation. So we conclude that
these ratios can be readily obtained by a multitude of natural
choices of the parameters $\rho$, $\alpha_1$, and $\alpha_2$
everywhere in the overall allowed parameter space of the model.

\section{Conclusions} \label{con}

We performed an analysis of the parameter space of the CMSSM with
$\mu>0$ supplemented by a generalized asymptotic Yukawa coupling
quasi-unification condition, which is implied by the SUSY GUT
constructed in Ref.~\cite{qcdm} and allows an experimentally
acceptable $b$-quark mass. We imposed a number of cosmological and
phenomenological constraints which originate from the CDM
abundance in the universe, $B$ physics ($b \rightarrow s\gamma$,
$B_s\to \mu^+\mu^-$, and $B_u\to\tau\nu$), and the mass $m_h$ of
the lightest neutral CP-even Higgs boson. We found that, in
contrast to previous results based on a more restrictive Yukawa
quasi-unification condition, the lightest neutralino can act as a
CDM candidate in a relatively wide range of parameters. In
particular, the upper bound from CDM considerations on the
lightest neutralino relic abundance, which is drastically reduced
mainly by $H$-pole enhanced stau-antistau coannihilation
processes, is compatible with the recent data on the branching
ratio of $B_s\to\mu^+\mu^-$ in this range of parameters. Also,
values of $m_h\simeq(125-126)~{\rm GeV}$, which are favored by
LHC, can be easily accommodated. The mass of the lightest
neutralino, though, comes out to be large ($\sim 1~{\rm TeV}$),
which makes its direct detectability very difficult and the
sparticle spectrum very heavy.

The fact that, in our model, $\Mg$, $m_0$, and $\mu$ generally
turn out to be of the order of a few $\TeV$ puts under some stress
the naturalness of the radiative electroweak symmetry breaking.
This is, though, a general problem of the CMSSM especially in view
of the recent data on the branching ratio of $B_s\to\mu^+\mu^-$
and the Higgs mass $m_h$ as noted in \cref{CmssmLhc, Antuch}.
Indeed, these data not only strongly restrict the parameter space
of the CMSSM so as to yield very heavy sparticle masses, but also
make the electroweak symmetry breaking less natural.

\acknowledgments We thank B.C.~Allanach, A.~Dedes, A.B.~Lahanas,
J.~Rizos, V.C.~Spanos, and N.~Tetradis for useful discussions.
This work was supported by the European Union under the Marie
Curie Initial Training Network `UNILHC' PITN-GA-2009-237920 and
the Greek Ministry of Education, Lifelong Learning and Religious
Affairs and the Operational Program: Education and Lifelong
Learning `HERACLITOS II'.

\def\ijmp#1#2#3{{Int. Jour. Mod. Phys.}
{\bf #1},~#3~(#2)}
\def\plb#1#2#3{{Phys. Lett. B }{\bf #1},~#3~(#2)}
\def\zpc#1#2#3{{Z. Phys. C }{\bf #1},~#3~(#2)}
\def\prl#1#2#3{{Phys. Rev. Lett.}
{\bf #1},~#3~(#2)}
\def\rmp#1#2#3{{Rev. Mod. Phys.}
{\bf #1},~#3~(#2)}
\def\prep#1#2#3{{Phys. Rep. }{\bf #1},~#3~(#2)}
\def\prd#1#2#3{{Phys. Rev. D }{\bf #1},~#3~(#2)}
\def\npb#1#2#3{{Nucl. Phys. }{\bf B#1},~#3~(#2)}
\def\npps#1#2#3{{Nucl. Phys. B (Proc. Sup.)}
{\bf #1},~#3~(#2)}
\def\mpl#1#2#3{{Mod. Phys. Lett.}
{\bf #1},~#3~(#2)}
\def\arnps#1#2#3{{Annu. Rev. Nucl. Part. Sci.}
{\bf #1},~#3~(#2)}
\def\sjnp#1#2#3{{Sov. J. Nucl. Phys.}
{\bf #1},~#3~(#2)}
\def\jetp#1#2#3{{JETP Lett. }{\bf #1},~#3~(#2)}
\def\app#1#2#3{{Acta Phys. Polon.}
{\bf #1},~#3~(#2)}
\def\rnc#1#2#3{{Riv. Nuovo Cim.}
{\bf #1},~#3~(#2)}
\def\ap#1#2#3{{Ann. Phys. }{\bf #1},~#3~(#2)}
\def\ptp#1#2#3{{Prog. Theor. Phys.}
{\bf #1},~#3~(#2)}
\def\apjl#1#2#3{{Astrophys. J. Lett.}
{\bf #1},~#3~(#2)}
\def\n#1#2#3{{Nature }{\bf #1},~#3~(#2)}
\def\apj#1#2#3{{Astrophys. J.}
{\bf #1},~#3~(#2)}
\def\anj#1#2#3{{Astron. J. }{\bf #1},~#3~(#2)}
\def\mnras#1#2#3{{MNRAS }{\bf #1},~#3~(#2)}
\def\grg#1#2#3{{Gen. Rel. Grav.}
{\bf #1},~#3~(#2)}
\def\s#1#2#3{{Science }{\bf #1},~#3~(#2)}
\def\baas#1#2#3{{Bull. Am. Astron. Soc.}
{\bf #1},~#3~(#2)}
\def\ibid#1#2#3{{\it ibid. }{\bf #1},~#3~(#2)}
\def\cpc#1#2#3{{Comput. Phys. Commun.}
{\bf #1},~#3~(#2)}
\def\astp#1#2#3{{Astropart. Phys.}
{\bf #1},~#3~(#2)}
\def\epjc#1#2#3{{Eur. Phys. J. C}
{\bf #1},~#3~(#2)}
\def\nima#1#2#3{{Nucl. Instrum. Meth. A}
{\bf #1},~#3~(#2)}
\def\jhep#1#2#3{{J. High Energy Phys.}
{\bf #1},~#3~(#2)}
\def\jcap#1#2#3{{J. Cosmol. Astropart. Phys.}
{\bf #1},~#3~(#2)}
\def\apjs#1#2#3{{Astrophys. J. Suppl.}
{\bf #1},~#3~(#2)}

\bibliographystyle{aipprocl}

\end{document}